\documentclass[a4paper,fleqn]{cas-dc}

\usepackage{amsmath,amssymb,amsfonts}
\usepackage{algorithmic}
\usepackage{graphicx}
\usepackage{textcomp}
\usepackage{caption}
\usepackage{wasysym}
\usepackage{float}
\usepackage[sort,numbers]{natbib}
\usepackage{multirow}
\usepackage{nomencl}
\usepackage{textcomp}
\usepackage{array}
\usepackage{wrapfig}
\usepackage{tabularx}
\usepackage{ulem}
\usepackage{etoolbox}

\renewcommand\nomgroup[1]{%
  \item[\bfseries
  \ifstrequal{#1}{A}{Abbreviations}{%
  \ifstrequal{#1}{B}{Terminology Definitions}{%
  \ifstrequal{#1}{C}{Symbols}{}}}%
]}
\newcommand{\nomunit}[1]{%
\renewcommand{\nomentryend}{\hspace*{\fill}#1}}

\makenomenclature

\def\tsc#1{\csdef{#1}{\textsc{\lowercase{#1}}\xspace}}
\tsc{WGM}
\tsc{QE}
\tsc{EP}
\tsc{PMS}
\tsc{BEC}
\tsc{DE}

\begin{document}

\let\WriteBookmarks\relax
\def\floatpagepagefraction{1}
\def\textpagefraction{.001}
\shorttitle{A Smart Meter Data-driven Distribution Utility Rate Model for Networks with Prosumers}
\shortauthors{A. Venkatraman et~al.}

\title [mode = title]{A Smart Meter Data-driven Distribution Utility Rate Model for Networks with Prosumers}

\author[1]{Athindra Venkatraman}[orcid=0000-0001-9857-2223]

\address[1]{Commonwealth Edison Company, 2 Lincoln Center, Oakbrook Terrace, IL 60181, USA}

\author[2]{Anupam A. Thatte}[orcid=0000-0002-0735-1597]

\address[2]{Midcontinent Independent System Operator, 720 City Center Drive, Carmel, IN 46032, USA}

\author[3]{Le Xie*}[orcid=0000-0002-9810-948X]
\address[3]{Department of Electrical and Computer Engineering, Texas A\&M University, College Station, TX 77843, USA}

\cortext[cor1]{Corresponding author: Le Xie \newline \textit{Disclaimer: The views expressed in this paper are solely those of the authors and do not necessarily represent those of ComEd or MISO.}}

\begin{abstract}
Distribution grids across the world are undergoing profound changes due to advances in energy technologies. Electrification of the transportation sector and the integration of Distributed Energy Resources (DERs), such as photo-voltaic panels and energy storage devices, have gained substantial momentum, especially at the grid edge. Transformation in the technological aspects of the grid could directly conflict with existing distribution utility retail tariff structures. We propose a smart meter data-driven rate model to recover distribution network-related charges, where the implementation of these grid-edge technologies is aligned with the interest of the various stakeholders in the electricity ecosystem. The model envisions a shift from charging end-users based on their KWh volumetric consumption, towards charging them a \textit{grid access fee} that approximates the impact of end-users' time-varying demand on their local distribution network. The proposed rate incorporates two cost metrics affecting distribution utilities (DUs), namely \textit{magnitude} and \textit{variability} of customer demand. The proposed rate can be applied to prosumers and conventional consumers without DERs. 

\end{abstract}



\begin{keywords}
Rate design; Distributed Energy Resources; Solar PV; Behind-The-Meter storage; Utility Cost Recovery 
\end{keywords}

\maketitle

\nomenclature[A]{DER}{Distributed Energy Resources}
\nomenclature[A]{EV}{Electric Vehicle}
\nomenclature[A]{PV}{Photo-Voltaic}
\nomenclature[A]{DU}{Distribution Utility: the Distribution Grid Operator}
\nomenclature[B]{Existing Mechanism:}{The volumetric \$/kWh rate model used in present rates}
\nomenclature[B]{Proposed Mechanism:}{The rate model based on demand metrics developed in this paper}
\nomenclature[C]{$X_i^t$}{Demand of Customer $i$ at time $t$ \nomunit{(kW)}}
\nomenclature[C]{$dX_i^t$}{Change (Variability) in Demand of Customer $i$ between intervals $t$ and $t-1$, i.e., $dX_i^t = X_i^t - X_i^{t-1}$\nomunit{(kW)}}
\nomenclature[C]{$W_i$}{Demand Magnitude Impact Factor for Customer $i$}
\nomenclature[C]{$V_i$}{Demand Variability Impact Factor for Customer $i$}
\nomenclature[C]{$\mu$}{Peak Demand Magnitude Indicator Function}
\nomenclature[C]{$\Pi_W$}{Allocation of Total Target Revenue for $W$}
\nomenclature[C]{$\Pi_V$}{Allocation of Total Target Revenue for $V$}
\nomenclature[C]{$k$}{Strictness of Peak Demand Percentile Cut-off}
\nomenclature[C]{$S^t$}{Total System Demand at time $t$ \nomunit{(kW)}}
\nomenclature[C]{$\beta^t$}{Total System Variability between intervals $t$ and $t-1$, i.e., $\beta^t = S^t - S^{t-1}$ \nomunit{(kW)}}
\nomenclature[C]{$S_\text{PeakTh}$}{Peak Demand Threshold}
\nomenclature[C]{$B_{i\text{ total}}^\text{old}$}{Total DU charge for Customer $i$ calculated under the existing mechanism \nomunit{(\$)}}
\nomenclature[C]{$B_{i\text{ total}}^\text{new}$}{Total DU charge for Customer $i$ calculated under the proposed mechanism \nomunit{(\$)}}
\nomenclature[C]{$\overline{\beta}$}{Mean of Total System Variability over all time-steps \nomunit{(kW)}}
\nomenclature[C]{$\overline{dX_i}$}{Mean Variability of Customer $i$ \nomunit{(kW)}}

\printnomenclature

\section{Introduction}
Distributed energy resources (DERs) have been integrated to the electric grid edge at an accelerated pace over the past decade. The levelized costs of photo-voltaic (PV) panels and energy storage have dropped significantly and are projected to continue this trend \cite{fu2018us}. Behind-The-Meter (BTM) technologies are estimated to make up over 50\% of the US energy storage market by 2021, with the deployed energy storage expected to reach 2 GW by then \cite{utilitydive}.

Although end-use demand is projected to increase in the next few decades both in the residential and commercial sectors, there is a significant projected reduction in energy intensity \cite{hostick2014projecting}. Further, projections indicate that the growth rate of electricity sales will be diminished due to the significant increase in generation from rooftop PV systems, from both residential and commercial buildings \cite{AEO2019}. The adoption of Electric Vehicles (EVs) is also on the rise, with the number of EVs on the road in the US reaching 1.1 million by the end of 2018 \cite{EVOutlook2019}. With increased installation of these technologies many consumers are turning into \textit{DER prosumers}, thus eroding the revenue stream of the utilities \cite{Kotilainen2019}.

The rise in DER penetration in markets around the globe makes the following question extremely relevant - are the existing distribution utility rate models poised to handle the accelerated pace of DER deployment at the grid edge? This study addresses customer rate models of Distribution Utilities (DUs) i.e., how they recover their costs from customers. DUs need to be compensated for their investments and the grid maintenance costs they incur to ensure reliable power supply to all customers. Their compensation is akin to a toll fee for using the DU’s grid infrastructure. We focus on the proper determination and allocation of \textit{grid access} costs to retail customers, and do not deal with the variable production costs for delivered power (the \textquotesingle Energy Charge\textquotesingle ) \footnote{\label{footnote1}Here we consider a scenario in which the variable production costs (Energy Charges) are passed through to the customer, and these are separate from the proposed grid-access fee. \cite{MIT_UF} states that "a comprehensive system of prices and charges consists of four core elements: (1) a price for electric energy; (2) prices or charges for other energy-related services, such as operating reserves or firm capacity; (3) charges for network-related services; and (4) charges to recover policy costs." Our paper focuses on (3) network-related services.}.

Existing DU charges are a combination of a small fixed charge, coupled with a larger volumetric \cent/kWh charge. The dotted line in Fig. \ref{fig:tducharge} represents the cost curve to consumers of this rate structure \cite{pennstate}. The revenue earned by DUs is directly proportional to the volume of electricity in kWh that is consumed by the end-users. This rate structure incentivizes DUs to maximize sales and makes them dependent on the volumetric charge for the bulk of their revenue \cite{faruqui2016curating}. With increasing deployment of grid-edge DERs the current rate design could be insufficient since it does not fully account for the rising fixed costs faced by the DUs \cite{mckinsey2019}. Grid-edge DERs pose a threat to the revenue stream of utility companies in a few different ways. First, the increase in solar PV penetration directly results in reduction of the kWh demand from the grid. This reduction lowers the customer's utility bill, even though the utility offers the service of access to the grid at all times, which the PV customer will require when sunlight is not available. Second, the expansion of participation in net metering has resulted in utilities providing financial compensation for injections of PV power to the grid \cite{Lawson2019}.

Costello \cite{costello2015major} argues that net metering is inefficient and results in cross-subsidy, and Brown et al. \cite{BrownSappingtonNM1, BrownSappingtonNM2} show that net metering is often not optimal. Gautier et al. \cite{GautierProsumer} show that net metering may lead to an uneconomic level of DER prosumption, with the lower bills for the DER prosumers compensated by higher bills for regular non-DER consumers. This form of cross-subsidy could lead to the utility death spiral \cite{BORENSTEIN_DEATHSPIRAL}.

Various case studies have been devoted to examining the impact of high DER penetration on the distribution grid and related technological, economic, and regulatory aspects. The resulting need for utilities to update their business models, and a rebalancing of costs on the electricity value chain from the grid side to behind the meter has been discussed in Woodhouse et al. \cite{innovation} and Smith et al. \cite{rebalancing} respectively. Baak \cite{DER_regulation1} and Alvarez Pelegry \cite{spanishreg} explore the regulatory framework in different parts of the world, and the restructuring that may be required to enable an accelerated transformation towards grid modernization, while Gellings \cite{noneedreg} argues that the existing regulatory measures may be adequate to accommodate even a transformed grid. Laws et al. \cite{laws2017utility} indicates that residential PV penetration could reach a substantial number over the next decade. But, they argue that utilities have ample time to change their business model in order to avoid the death spiral. Darghouth et al. \cite{darghouth2016net} shows how various rate design choices can impact the long term cumulative distributed PV deployment.

Gautier et al. \cite{GautierProsumer} build a model to compare net metering and net purchasing, concluding that net purchasing is a better rate model for a network with prosumers. Cambini et al. \cite{Cambini_Multipart} observe that net metering remains the mechanism most compatible with existing metering technologies, and propose a multi-part tariff for DER prosumers, with explicit components for connection costs, distribution costs, and energy losses. However, the proliferation of Advanced Metering Infrastructure (AMI) technologies provides an enabling platform for retail rate innovations that could improve upon the current volumetric delivery rate structure.

In Bharatkumar \cite{DNUOS}, a Distribution Network Use-of-System (DNUoS) charge has been proposed, which aids in the accurate recovery of distribution utility costs, by capturing the contribution of each user on the network to the system's costs. This study applies a similar line of thought, by billing customers based on a novel measure of their individual impact on distribution system costs.

The difference in our approach is the consideration of the granular inter-temporal impact factor, i.e., \textit{when} the customers impact the grid. This approach is akin to the idea of coincident-peak capacity charges, discussed in \cite{MIT_UF,Abdelmotteleb,Baldick2018,PASSEY2017642,Lazar2020}. We aim to use higher granularity AMI data to more accurately allocate the cost incurred by distribution utilities to their customers. The overall goal is to design rates that are more reflective of the cost drivers, thereby better aligning the incentives for both infrastructure owners and technological innovation at the grid edge.

Both Gautier et al. \cite{GautierProsumer} and Cambini et al. \cite{Cambini_Multipart} base their respective models on the assumption of a fixed retail volumetric kWh unit price for energy. Our model is a departure from the conventional volumetric unit price for network-related costs, and proposes a data-driven approach to approximately estimate the contribution of customers to system costs. 
Our model applies to all end-users, including DER prosumers and non-DER consumers, unlike the model proposed by Cambini et al. \cite{Cambini_Multipart}, which only applies to DER prosumers.

In view of the above, the key contributions of this research are as follows: 
\begin{itemize}
  \item We propose new metrics to quantify the impact of customers on the grid, based on demand magnitude and variability.
  \item We introduce an  alternative billing determinant that calculates DU grid-access charges for customers based on the proposed demand metrics. 
  \item We conduct realistic numerical case studies to illustrate the efficacy of the proposed rate structure, especially for customers solar PV and electric vehicles. 
\item We simulate and analyze the deployment of battery storage in the new rate structure to illustrate its major impact on the grid and on the customers.
\end{itemize}

The rest of the article is organized as follows: Section \ref{method-section} highlights the drawbacks of the existing distribution utility rate model, and describes the design details and mathematical formulation of the proposed rate model. Section \ref{results-section} is a critical comparison of the existing and proposed distribution utility rate models, supported by a case study using real residential customer data. Section \ref{conclusions-section} summarizes the key lessons and the most significant policy implications of the proposed distribution utility rate model.

\section{Methodology and Data}\label{method-section}

\subsection{Proposed Distribution Utility Rate Model}

\begin{figure}
	\centering
	\includegraphics[width=0.9\linewidth]{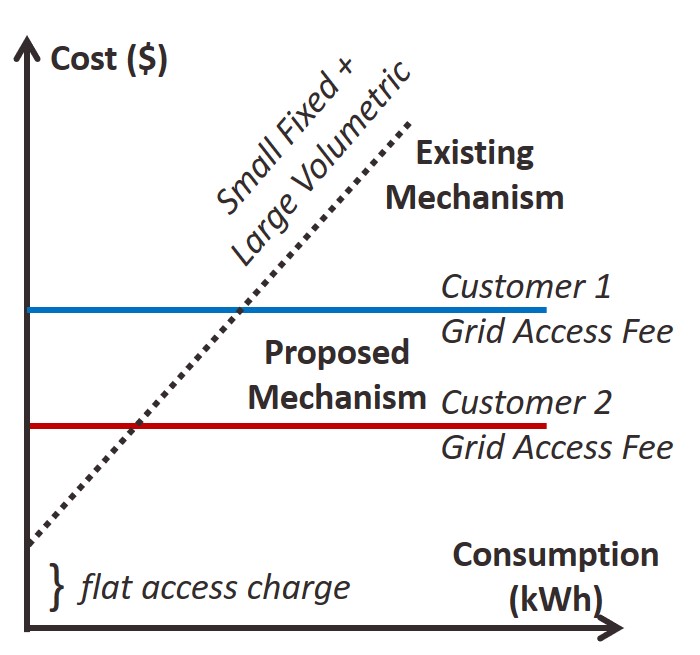}
	\caption{Existing vs. Proposed DU charges: Consumer Cost Curve}
	\label{fig:tducharge}
\end{figure}

The proposed DU charge features the introduction of a \textbf{grid-access fee}, replacing the existing distribution utility charges, which are typically structured as a small flat access charge combined with a large volumetric (\$/kWh) charge (see footnote \ref{footnote1}). Fig. \ref{fig:tducharge} presents a graphical comparison of the existing and proposed utility rate models. The uniqueness of this idea lies in how these grid-access fees would be calculated, by taking into account some key parameters that define the \textit{impact} of customers to the grid. This impact is quantified through a combination of cost metrics called Grid Impact Factors. 

This concept is analogous to an insurance rate model or a credit score, where each customer’s rate or credit limit is considered to accurately reflect the \textit{risk level} taken up by the insurance company or bank by entering into business with said customer. An example of providing incentives in wholesale electricity markets is FERC Order 755 \cite{ferc755}. Prior to the adoption of the order, most ISO markets in the US had a single capacity payment for regulation. Payments from ISOs to generators compensating them for frequency regulation service were not tied to the performance of the resource. As a result of this rule a two-part payment was enacted, which added a mileage-based component that accounts for the performance of the resource, in addition to the capacity based payment.

In the following case study, we describe an idealized setup where the grid-access fees are tailored to individual customers. Recognizing the limitations of the approach, we propose a practical implementation mechanism in Section \ref{implementation}.

As is evident from Fig. \ref{fig:tducharge}, the DU does not have revenue assurance i.e., they have to hope that consumers use more kWh, thus driving up their revenue. However, under the proposed approach the DU has a steady and assured income from each consumer via the fixed Grid Access Fees. In the figure, Customer 2 has a lower grid impact than Customer 1, and thus is charged a lower Grid Access Fee.

\subsection{Cost Drivers for Distribution Utilities}
In the context of a distribution grid, the installed capacity of the system is a key parameter; it determines how much load can be served. Depending on changing load patterns this limit also dictates the need for capital investment. System capacity requirements are directly dependent on the system Peak Demand to be supplied to the customers. Thus, the "Peak Demand Time Slots" are a critical time for the system. To account for this, a Demand Magnitude Impact Factor $W$ is introduced, that measures the demand impact factor of each home during the peak demand time slots.
\begin{equation}
    \begin{aligned}
       &\text{Demand Impact Factor of Home $i$ }    (W_i) \\ &
       = \text{Total Demand of Home } i \text{ during Peak Slots}
     \end{aligned}
\end{equation}

Another key concern for the distribution grid is the health of the existing infrastructure, directly impacting the capital investment and maintenance costs that the DU incurs. McBee \cite{MCBEE7923954} describes how a high penetration of PV, EV charging, and energy storage causes a significant increase in transformer aging due to higher total harmonic distortion, and almost constant energy demand over long periods. Awadallah et. al. \cite{awadallah} quantify the harmonic distortion caused by PV on distribution transformers, and find that with high penetration of PV, transformer life expectancy could decrease by 8.3\%. Various studies have shown that for higher penetrations of PV in the distribution system, the number of transformer tap changes increase rapidly even for small levels of variability \cite{Dong, YAN2014185}, sometimes up to a sixfold increase on high variability days \cite{malaysia}. Dubey et al. \cite{DUBEY7264982} provide a comprehensive study on the impact of EV charging on residential distribution systems, where voltage variations at EV load locations is identified as a key issue. 

Thus, the health of grid infrastructure is correlated to its loading conditions and the fluctuations in demand. These fluctuations are measured using the Demand Variability Impact Factor $V$, which quantifies customer usage contribution to system variability by computing the normalized correlation between the variability of customers demand and the variability of total system demand. 

\begin{equation}
    \begin{aligned}
       &\text{Variability Impact Factor of Home $i$ } (V_i) \\ &
       = \text{Normalized Correlation between} \\& \text{Variability of Home } i \text{ and Total System Variability}
     \end{aligned}
\end{equation}

The inclusion of the Variability Impact Factor $V$ is a novel concept. We build upon this concept to propose a compatible distribution utility rate model to reflect this time-varying impact on the network-related cost.

\subsection{Peak Demand Indicator Function} \label{peaktime}

To make the rate structure as flexible and general as possible, a Peak Indicator Function for Demand has been introduced. This function takes as input the present system conditions, the peak threshold for the system conditions, and a strictness parameter $k$, to deliver an indication of whether the system condition at that time $t$ is considered to be a peak slot or not. When the value of $k$ is very low, the Peak is considered based on a stringent cut-off, whereas if the value of $k$ is larger, the function also begins to consider those time slots where System Demand ($S^t$) is almost equal to the peak threshold, thus reducing the importance and emphasis placed on an inherently arbitrary definition of peak threshold.

Distribution grids have diverse load profiles, even between different feeders within the same DU service territory. The proposed mechanism provides the DU with the option to select peak thresholds and $k$ values that are best suited for their system conditions. Optimal peak thresholds can be determined through engineering analysis performed by the DU, with oversight of the regulator.

The Peak Indicator Function has been defined for Peak Demand Magnitude as $\mu$. This function is described below.

\subsubsection{Peak Demand Indicator Function $\mu$}
\begin{figure}
	\centering
	\includegraphics[width=\linewidth]{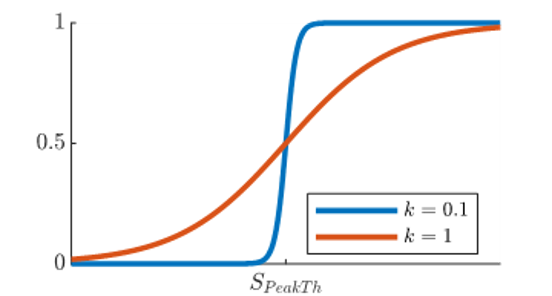}
	\caption{Peak Demand Indicator Function $\mu$}
	\label{fig:mu}
\end{figure}

\noindent This function is designed similar to a logistic function, and is centered around the System Peak Threshold value $S_\text{PeakTh}$. $S_\text{PeakTh}$ is calculated based on a percentile value that is set by the DU. If the peak threshold percentage is set as $15\%$, then $S_\text{PeakTh} = 85^{th}$ percentile of the system load curve. This means that a given time slot $t$ is defined as a peak demand time slot when $S^t \geq S_\text{PeakTh}$.
In essence, this function returns $1$ if it is a peak slot, and $0$ if not (Fig. \ref{fig:mu}). For a given time $t$, $\mu$ is described as follows:

\begin{equation}
   \mu^t = \frac{1}{1+e^{\frac{-(S^t - S_\text{PeakTh})}{k}}}
\end{equation}

\subsection{Calculating the Grid Impact Factors $W$ and $V$}
\noindent
Let $X_i^t =$ Demand of user $i$ at time $t$ \newline
$dX_i^t =$ Change in Demand (Variability) of user $i$ between time $t$ and $t-1$\newline
i.e., $dX_i^t = X_i^t - X_i^{t-1}$

\subsubsection{Demand Magnitude Impact Factor $W$}
Demand Magnitude Impact Factor for Customer $i$ ($W_i$) is the total demand of Customer $i$ during peak demand time slots (defined in Section \ref{peaktime}). The Peak Indicator Function $\mu$ is used to determine whether time-step $t$ is a peak or not.
\paragraph{Element-Wise Multiplication}
\begin{equation}
W^t_i = X_i^t \cdot \mu^t\\
W_i = \sum_t{W_i^t}
\end{equation}

\paragraph{Matrix Multiplication}
\begin{equation}
 W_{N \times 1}  = X_{N\times T} \cdot \mu_{T\times 1}\\
N \text{ homes}, T \text{ timesteps}\\
\end{equation}

\subsubsection{Demand Variability Impact Factor $V$}

The Demand Variability Impact Factor for customer $i$ ($V_i$) is the normalized correlation between the variability of customer $i$ ($dX_i$) and the variability of total system demand $\beta$.
\begin{equation}
\begin{aligned}[c]
V_i &= \frac{\sum_t(dX_i^t - \overline{dX_i})(\beta^t - \overline{\beta})}{\sqrt{\sum_t{(dX_i^t - \overline{dX_i})} \sum_t{(\beta^t - \overline{\beta})} }}
\end{aligned}
\end{equation}

\subsubsection{Relative Factor (\% allocation) for each Customer $i$}
\begin{equation}
\begin{aligned}[c]
W_{\text{share }i} = \frac{W_i}{\sum_{j=1}^N W_j}\\
\end{aligned}
\qquad \qquad
\begin{aligned}[c]
V_{\text{share }i} = \frac{V_i}{\sum_{j=1}^N V_j}\\
\end{aligned}
\end{equation}

\subsection{Calculating the Final Bills}
\noindent
Let $B_{i\text{ total}}^\text{old} \rightarrow$ Total DU charge of home $i$ calculated in the current method, and $B_{i\text{ total}}^\text{new} \rightarrow$ Total DU charge of home $i$ calculated in the proposed method.

To calculate DU charges for each home under the \textbf{current mechanism}, we consider a standard volumetric rate formula for DU charges defined below (5\cent \space per kWh) ($X_i^t > 0$):
\begin{equation}
    B_{i}^{t\text{ old}} = \$0.05\times X_i^t
\end{equation}

In case a home generates more than it consumes at any point in time, i.e., $X_i^t < 0$, the excess electricity is sold back to the grid at a discounted rate of 2\cent \space per kWh (Net Metering).
\begin{equation}
    B_{i}^{t\text{ old}} = -\$0.02\times X_i^t
\end{equation}

So, the total DU charges in the current mechanism for the full 2 year period is calculated as follows.

\begin{equation}
    B_{i\text{ total}}^\text{old} = \sum_t(B_i^{t\text{ old}})
\end{equation}

In the \textbf{proposed distribution utility rate model}, customers are charged a cumulative grid access charge (e.g. on a monthly basis), based on their Grid Impact Factors. This access charge is calculated by starting from the total target revenue for the DU, and dividing this total among customers based on their contribution to grid impact.

The access charge for customers remains the same during one assessment period, and is updated at the end of each assessment period based on usage data collected to-date.

It is assumed that the \$/kWh rate is derived from the total target revenue of the system, which is obtained as a result of the current rate case process.

For simplicity, the analysis operates under the assumption that the total target revenue is calculated for the full period of assessment. In our case study, this period of assessment is two years. Further, to make a fair and direct comparison of the current and proposed rate models, this total target revenue for the DU has been fixed as the $B_{i\text{ total}}^\text{old}$ value, i.e.,
\begin{equation} \label{eq:10}
    \sum_{i=1}^N B_{i \text{ total}}^\text{old} = \sum_{i=1}^N B_{i \text{ total}}^\text{new}
\end{equation}

The result is essentially a \textit{redistribution} of the same final cost among the customers. This assumption is a fair one to make because the total target revenue for the current mechanism is calculated through the rate case process, which is assumed to be an accurate reflection of system costs.

Since the proposed mechanism has to account for two contributing grid impact factors $W$ and $V$, the importance of these respective weighting factors are determined by the allocation percentage parameters $\Pi_V$ and $\Pi_W$ (also determined by the DU and regulator), defined as follows:

\noindent
$\Pi_W = \%$ allocation of Total Target Revenue for $W$
\newline
$\Pi_V = \%$ allocation of Total Target Revenue for $V$

And so, finally, the total DU charges for each home $i$ as per the proposed scheme is calculated as a linear combination of the weighting factors scaled with their respective allocation percentage parameters, as follows:
\begin{equation}
    B_{i\text{ total}}^\text{new} = W_{\text{share }i} \times \Pi_W + V_{\text{share }i} \times \Pi_V
\end{equation}

\subsection{Data and Case Study System Description}

The data used for the results discussed in Section \ref{results-section} is the instantaneous kW demand for 200 residential customers, measured at a resolution of 1-minute. The dataset spanning two years (from 01-01-2016 to 12-31-2017) was obtained from Pecan Street Dataport \cite{pecanstreet}.


\section{Results and Discussion} \label{results-section}

To thoroughly examine the effects of the proposed distribution utility rate model, we calculate the DU charges for each home in a system of 200 residential demand profiles, with 25\% penetration of EVs and PVs each, i.e., 50 EV homes and 50 PV homes among the 200 total homes. 

For the purpose of this example, we set the $S_\text{PeakTh}$ at the 75\textsuperscript{th} percentile of total system demand. Also, the \% allocations of Total Target Revenue $\Pi_W$ and $\Pi_V$ are set as 75\% and 25\% respectively.

\subsection{Comparing the Performance of Two Homes in the Proposed Mechanism}
\begin{figure*}
	\centering
	\includegraphics[width=\linewidth]{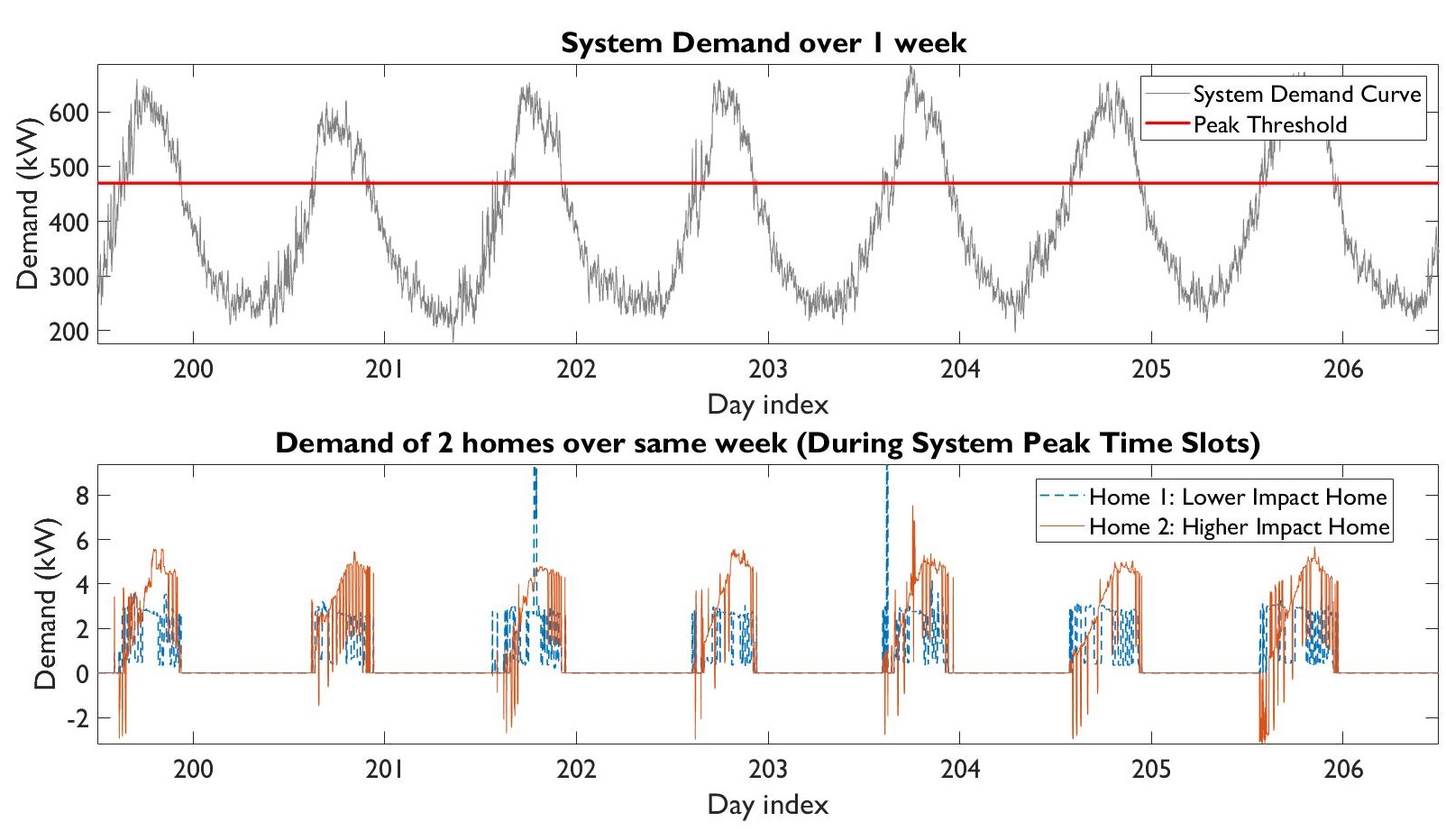}
	\caption{System Demand Curve and Comparison of Demand between two homes over one week during Peak Time Slots}
	\label{fig:winnerloser}
\end{figure*}

To illustrate the effects of the proposed scheme, we examine two homes which have a similar DU charge in the existing scheme but a significant difference in DU charges in the proposed scheme.

In Fig. \ref{fig:winnerloser} (top), the system demand curve has been plotted along with the Peak Threshold line (red), indicating which intervals are considered to be peak time slots. Fig. \ref{fig:winnerloser} (bottom) depicts the individual demand of the higher impact and lower impact homes during the system peak time slots, and is assumed to be zero for non-peak time slots.

Despite having a few spikes of demand, the demand of Home 1 during the peak time slots is, for the most part, less than that of Home 2. Furthermore, Home 2 has a negative demand for several periods each day i.e., it is generating more power than it consumes, indicating that it is a solar PV home. The fact that this home is a \textit{higher impact home} can be attributed to the benefit given to solar PV homes in the existing mechanism due to net metering. In the proposed mechanism, such demand variability is penalized through the $V$ parameter.

\subsection{Comparing the Current and Proposed Distribution Utility Rate Models}

\begin{figure*}
	\centering
	\captionsetup{justification=centering}
	\includegraphics[width=\linewidth]{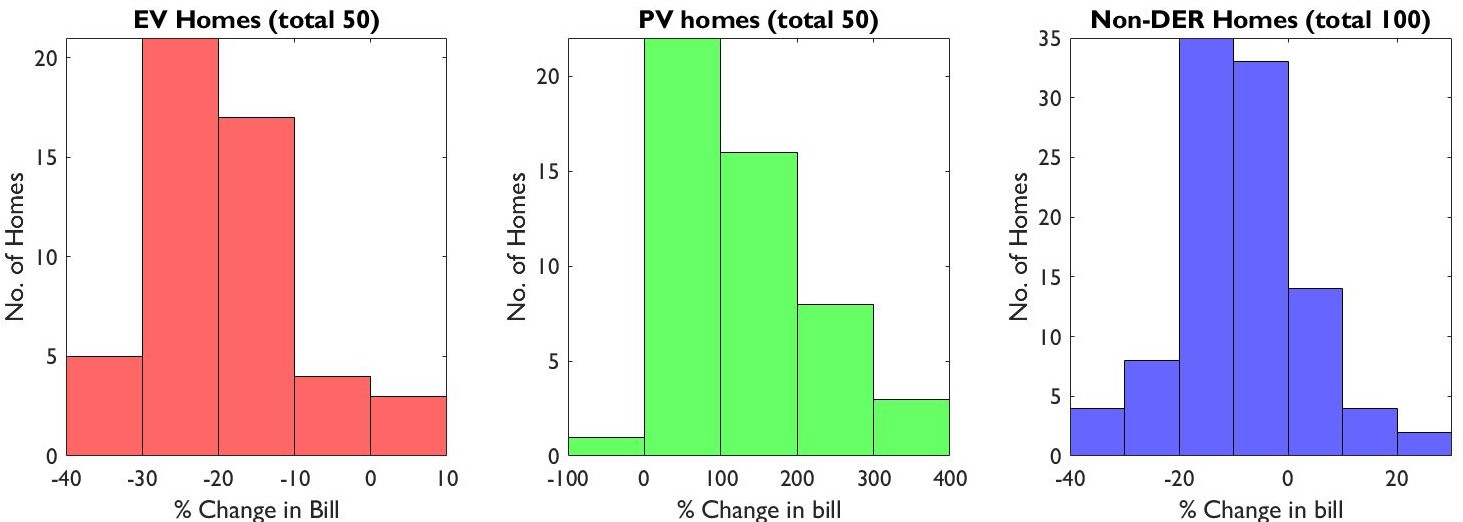}
	\caption{Comparison of Proposed and Existing Distribution Utility Rate Models for Homes}
	\label{fig:hist_oldvnewbill}
\end{figure*}

Fig. \ref{fig:hist_oldvnewbill} describes the effect of the proposed rate model for each subset of homes.
This effect is quantified by evaluating the \textbf{percentage change} between the proposed bill and the current bill, i.e., $B_{i\text{ total}}^\text{new} - B_{i\text{ total}}^\text{old}$ for each home. The distribution of this range has been plotted, categorized based on the type of home: EV Homes, PV Homes, and non-DER Homes.

In the case of EV homes, most homes have a negative \% change of $B_{i\text{ total}}^\text{new} - B_{i\text{ total}}^\text{old}$. Thus, almost all homes have a lower DU charge in the proposed mechanism than they do in the current mechanism. As a result, it seems that the proposed rate model is \textit{favorable for EVs}. This observation follows intuition, because under the current rate model, all that matters for billing is the kWh volumetric consumption by the home. Whereas in the proposed billing algorithm, the \textit{impact} of the user is calculated during the peak time slots of demand, where the distribution system is under the most stress. Thus, under the proposed rate model, there is great potential for \textit{smart scheduling} of EV charging during the non-peak periods, which could lead to significant savings for those homes. As a result, the interests of both the DU and the user are aligned.

When we observe the trend for PV homes, most homes have a positive \% change of $B_{i\text{ total}}^\text{new} - B_{i\text{ total}}^\text{old}$, which means that almost all PV homes have a significant increase in their DU charge when evaluated under the proposed mechanism. While this observation seems to suggest that the proposed mechanism is \textit{unfavorable to PV homes}, it can be argued that the proposed mechanism is capturing the actual costs of PV that were previously (unfairly) being borne by non-PV homes. Although the kWh volume of consumption for PV homes is relatively lower, the sudden ramping of PV during the late evening causes significant strain on the distribution grid. This aspect is captured in the proposed billing scheme through the Variability impact factor $V$. 

Let us now consider the case of non-DER homes. Most homes have a negative percentage change value for $B_{i\text{ total}}^\text{new} - B_{i\text{ total}}^\text{old}$. More specifically, of the 100 non-DER homes, over 80 have a negative $B_{i\text{ total}}^\text{new} - B_{i\text{ total}}^\text{old}$, with almost 70 homes having a slightly negative change (0-20\% reduction in the bill). This indicates that the proposed rate model is benefiting most non-DER homes. This addresses one of the key drawbacks of the existing billing scheme, where, in many cases, costs incurred by the DUs in their PV-incentive programs such as net metering or other subsidies would be recovered from the non-PV customers via an increase in access charges. With the proposed mechanism, the trend of penalizing non-PV customers is reversed, bringing the distribution of DU charges back to balance.

\subsection{The Effect of DER Penetration on DU charges Calculated under the Proposed Mechanism}

\begin{figure*}
	\centering
	\includegraphics[width=\linewidth]{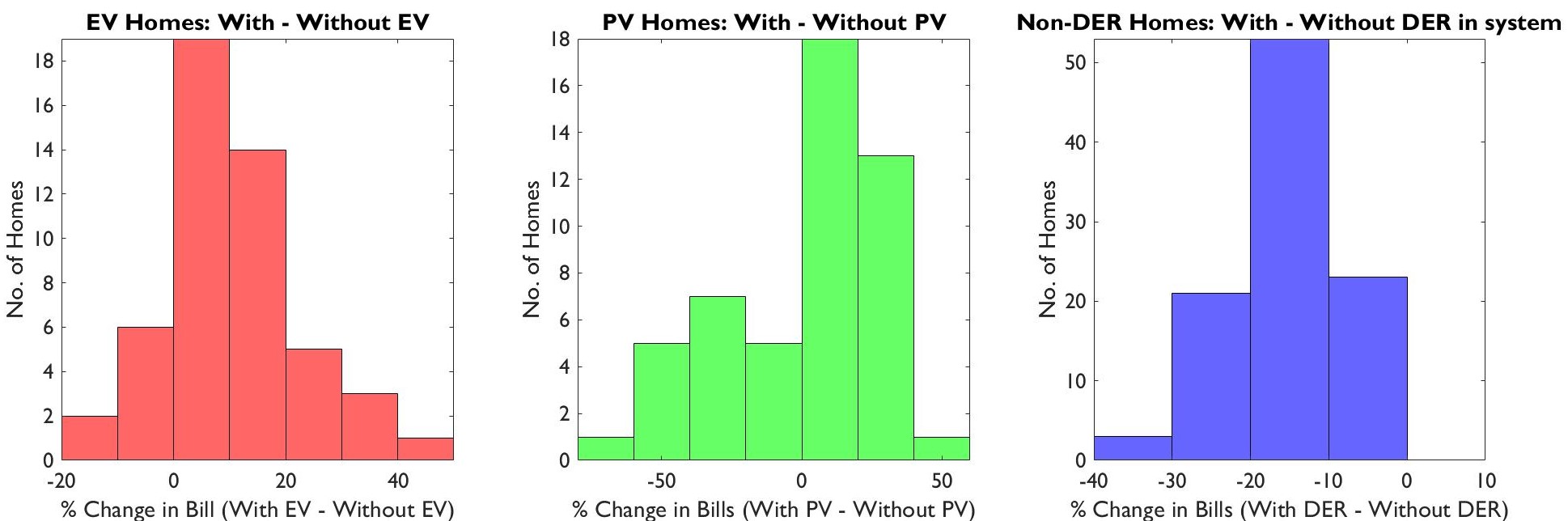}
	\caption{\% Change in DU charges before and after 25\% DER Penetration (Calculated under the Proposed Mechanism)}
	\label{fig:hist_DERvsNoDER}
\end{figure*}

Fig. \ref{fig:hist_DERvsNoDER} describes the effect of penetration of individual DERs (EV and PV) on each subset of homes. In the default system, there is a DER penetration of 25\% EV and 25\% PV (50 homes each). In the system without EV, the DER penetration is 0\% EV (0 homes) and 25\% PV (50 homes). Similarly, in the system without PV, the DER penetration is 25\% EV (50 EV homes) and 0\% PV (0 PV Homes). In the system without DERs, the DER penetration is 0\%, i.e., 0\% EV and 0\% PV. The left figure compares the DU charges of the EV Homes calculated in the default system \textit{vs} the system without EVs. The middle figure compares DU charges of PV homes calculated in the default system \textit{vs} the system without PV generation. The right figure shows the effect on DU charges of non-DER homes due to DER penetration in the system, by comparing the DU charges calculated in the default system \textit{vs} the system with 0\% DER penetration.

When considering the effect of EV penetration on EV homes, it is observed that most homes have a positive \% change between with and without EV cases, thus following the expected trend of having higher electricity bills due to the presence of an EV. 

With PV however, the story is different. Some PV homes seem to benefit with the introduction of PV (around 30 homes), but the rest have a higher DU charge with the introduction of PV. One factor could be explained by the variability index $V$ accounting for 25\% of the total revenue, and that the PV homes have the highest variability impact factors. Another issue could be that PVs are pulling down the system conditions below peak threshold when the sun is shining, and shifting peak slots to different times. Thus, the application of solar + storage technology combined with smart scheduling for maximizing usage during system non-peak conditions could be the optimal strategy in the proposed billing scheme. This has been explored in the case study discussed in Section \ref{battery_analysis}. 

\begin{figure*}
	\centering
	\includegraphics[width=\linewidth]{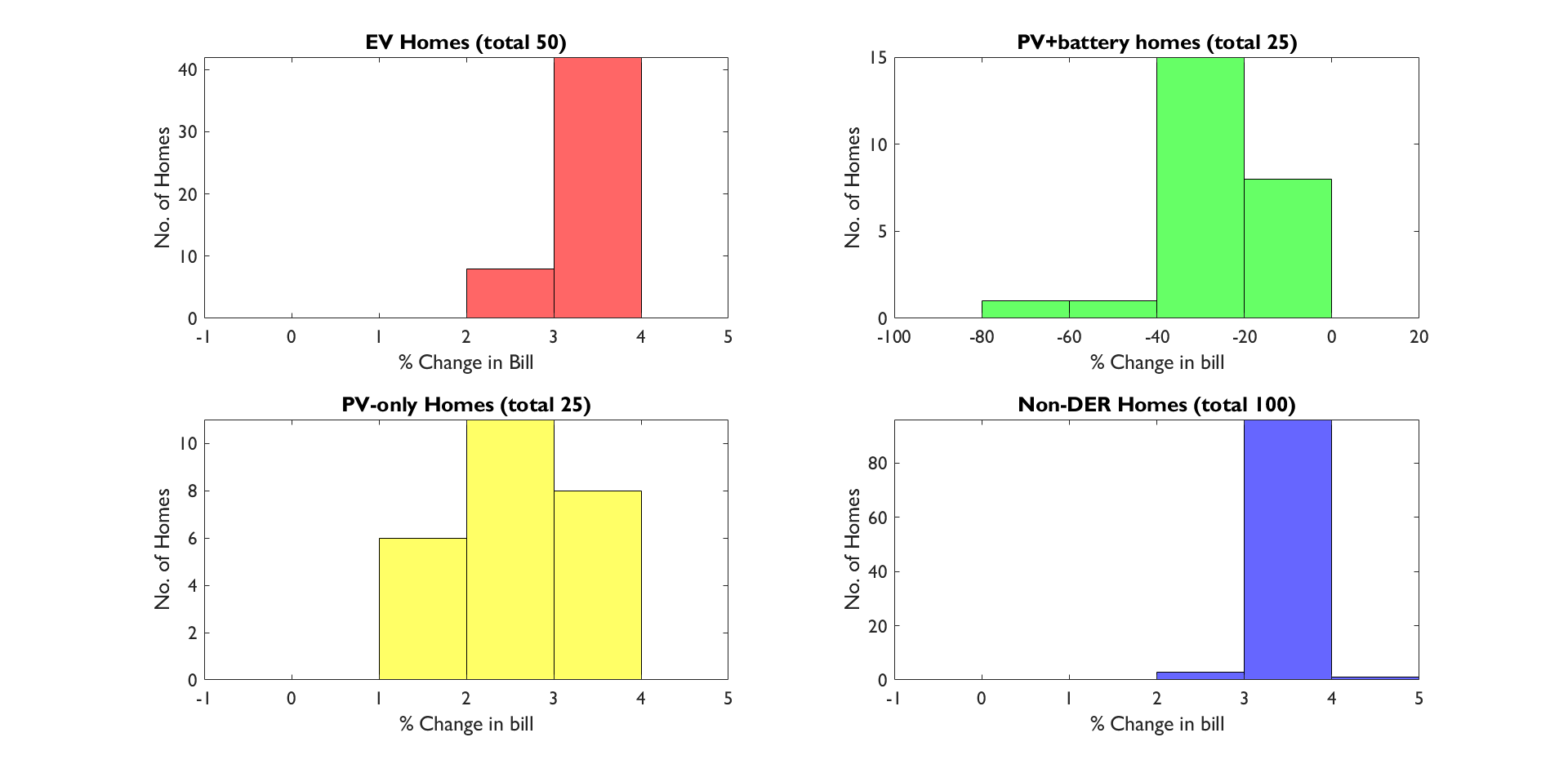}
	\caption{\% Change in DU charges (Proposed Mechanism) after introducing Battery Storage in 25 PV Homes}
	\label{fig:hist_batterypctchange}
\end{figure*}

Looking at the effect of DER penetration on non-DER homes, it is noted that every single non-DER home has seen a reduction in their DU charges due to the penetration of DER. While this seems like the proposed mechanism rewards customers for not investing in DER, it is more accurate to view this as evidence that a fair cost recovery from DER homes is happening because of DER homes having an increase in their grid impact, due to the penetration of DERs. 

\subsection{Effect of Battery Storage on DU charges calculated under the Proposed Mechanism}\label{battery_analysis}

Fig. \ref{fig:hist_batterypctchange} shows the effect of penetration of battery storage in the system on the DU charges calculated under the proposed rate mechanism. In this case study, half of the PV homes (25 out of 50) are given a battery storage unit, that operates under a brute force algorithm, charging during typical non-peak hours (1 am to 3 am), and discharging during typical peak hours (5 pm to 7 pm), with a rate of 2 kW for both charge and discharge cycles. Essentially, this is meant to reduce the impact on the grid by discharging during peak time slots, and charging during non-peak time slots.

Fig. \ref{fig:hist_batterypctchange} (top-right) shows that every single 'PV+battery' home has experienced a reduction in DU charge due to the introduction of battery storage. This reduction has been observed even though a brute force charging-discharging schedule was implemented. This result could be further improved if the battery storage devices are operated under a smart-scheduling algorithm, that not only reduces impact during peak time slots, but also counteracts spikes in the variability of the system, thus earning \textit{rewards} for positive contributions to grid conditions.

The other three sub-figures in Fig. \ref{fig:hist_batterypctchange} show the effects of the introduction of battery systems in 25 PV homes on the other categories of homes. Homes in all of these categories see minor increases in their bills, so it could be argued that the proposed mechanism provides the most rewards for customers having PV + battery storage, who are more likely to be wealthier customers, at the expense of non-DER customers, who may be less affluent and cannot afford PVs and battery storage. However, when comparing the DU charges of these non-DER customers under the proposed and current schemes
, it is clear that these homes will still be better off than they are under the current scheme.

\subsection{Effect of Customer Aggregation on DU charges calculated under the Proposed Mechanism} \label{aggregationsection}

\begin{figure*}
	\centering
	\includegraphics[width=\linewidth]{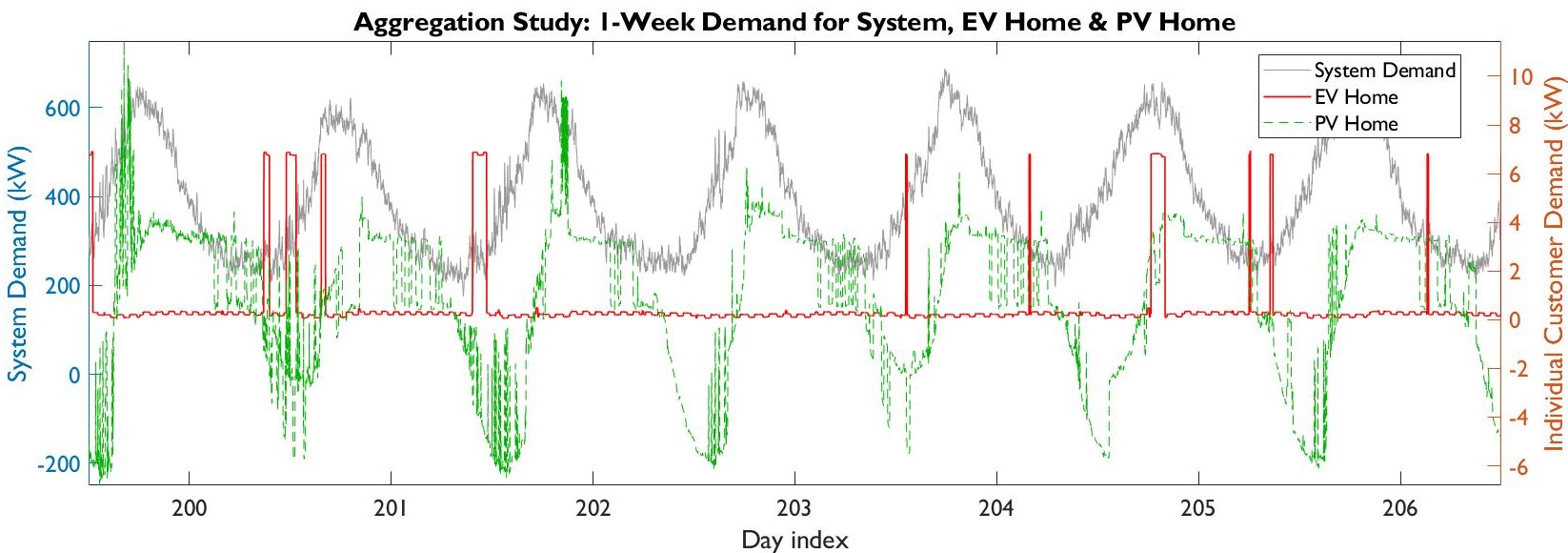}
	\caption{Case Study: Aggregating a low-$V_i$ EV customer and a high-$V_i$ PV customer}
	\label{fig:aggregation}
\end{figure*}

Consider the case of customer aggregation, where a group of customers come together to form an aggregated customer group, which is essentially treated as a singular customer entity by the DU. It is desirable for customers within this group to complement each other's variability $V_i^t$ such that there is a reduction in the net variability impact of the aggregated customer group. If net variability impact of a group reduces, the total impact on the system reduces. Such aggregated customer groups could potentially be governed by an internal smart control algorithm such that the group minimizes their net impact on the system. An efficient rate model should reflect rewards for a desirable reduction in system impact.

In the existing rate model, there would be no change in the total DU charge of these aggregated customers, since the total kWh consumption remains the same. However, the proposed mechanism considers variability of power demand as a key parameter for evaluating the impact of customers. Thus, such aggregated customer groups will have a net reduction in aggregate customer bill, compared to the sum of DU charges of the same customers treated as individuals.

Fig. \ref{fig:aggregation} is a case study performed with two customers to illustrate the effect of aggregation on DU charges calculated in the proposed mechanism. Here, a low-$V_i$ EV customer is aggregated with a high-$V_i$ PV customer. No smart control algorithm has been implemented to change the behavior of either customer; this case study is conducted to illustrate the potential of customer aggregation for DU charge reduction. Fig. \ref{fig:aggregation} shows the demands of the system, the EV customer, and the PV customer over one week. These homes oppose each other's variability quite regularly, especially when the EV customer's EV is charging and the PV customer's solar panel is generating power (note the multiple 6.6 kW spikes seen in the EV customer's demand curve and the intermittent spiking of the PV customer's demand curve). We expect a net reduction in variability impact when these customers are aggregated, thus resulting in a lower DU charge for the aggregated customer group compared to the sum of the two individual bills. The results of this case study are shown in Table \ref{table}.

We observe that customer aggregation could have the desired effect of further incentivizing variability impact reductions, which can be harnessed by smart control algorithms governing the behavior of customers within such aggregated customer groups.

\begin{table*}
\begin{tabular}{m{15em}ccc}
\multirow{2}{*}{\textbf{Evaluation Metric}}                 & \multicolumn{2}{c}{\textbf{Individual Customers}} & \multirow{2}{*}{\textbf{If same customers were Aggregated}}                                               \\
                                                            & \textbf{PV Customer}    & \textbf{EV Customer}    &                                                                                                           \\
\% Share of Variability Impact using Normalized Correlation & 2.11\%                  & 0.27\%                  & 1.85\%                                                                                                    \\
\multirow{2}{*}{Total Bill over full 2-year period (\$)}    & \$1913                  & \$622                   & \multirow{2}{*}{\begin{tabular}[c]{@{}c@{}}\$2234\\ \$301 (11.87\%) saved through aggregation\end{tabular}} \\
                                                            & \multicolumn{2}{c}{Total = \$2535}                &                                                                                                          
\end{tabular}
\caption{Case Study: Comparing Impact and Total DU charges before and after Aggregation}
\label{table}
\end{table*}

\subsection{Pros and Cons of the Proposed Mechanism}

\subsubsection{Pros}
\paragraph{\textbf{Revenue Decoupling}}

The mechanism introduced in this work effectively decouples utility revenue and customer bills from volumetric consumption. Revenue decoupling is important to the long-term stability of utility revenues, since due to the growing penetration of DERs, volumetric based revenue could decline in the future.

\paragraph{\textbf{Recovers Utility Costs Accurately and Effectively}}

The proposed mechanism is more representative of the actual costs inflicted upon the distribution grid by the customers, due to the usage of kW rather than kWh as a defining metric. The major driver for investment costs in equipment is the consumer demand during peak periods. Thus, the proposed approach provides better alignment between the revenue and costs as compared to the volumetric charge. The introduction of 'Variability' is also a novel approach. The variable nature of renewable resources adversely impacts the efficient operation of the grid and as such should be accounted for in the cost recovery mechanism.

\paragraph{\textbf{Utility Revenue Targets are Assured to Be Met}}

There is a key and prominent distinction between the proposed mechanism and the current model: rather than expecting a total revenue for the DU depending on several variables, the proposed mechanism offers the DU the opportunity to ensure a stable and assured revenue. The total target revenue is first set, then the proposed mechanism allocates the costs to all customers. Another advantage is that this form of rate-making could require less frequent rate cases, which is a time-consuming and expensive process.

\paragraph{\textbf{Reduces Unfair Cross-Subsidy}}
Both the current volumetric charge and the net-metering policies result in DUs over-recovering costs from non-PV customers while under-recovering them from PV customers. Further, there is a high likelihood that non-PV customers fall in the low-income category, while PV customers fall in the high-income category, as observed by De Groote et al. \cite{DEGROOTE201645}. Thus the proposed mechanism mitigates against the existing unfair and regressive cross-subsidy.
Further, the proposed approach is consistent for all types of DERs. This consistency is relevant to incentivize technologies such as energy storage.

\paragraph{\textbf{Retains Efficiency Incentive}}
Under the current volumetric mechanism increasing efficiency reduces electricity sales and therefore profits \cite{jensen2007aligning}. The proposed rate model retains the incentive for the DU to be efficient. Since the total revenue target is controlled under this structure, the DU is incentivized to take action to improve system efficiency to get higher profits.

\paragraph{\textbf{Rewards Smart Customer Aggregation}}

The proposed rate structure could incentivize aggregation of customers with consumption patterns that are negatively correlated with each other. As illustrated in Section \ref{aggregationsection} and Fig. \ref{fig:aggregation}, such groups of customers could reduce their aggregate impact on the grid, thus reducing their DU charge when part of a group, compared to their DU charge when considered individually.

Smarter scheduling and real-time adaptive consumption patterns could be leveraged in such aggregation mechanisms to negate the spikes introduced by other customers in the group. This also opens up the market for customer aggregation as an electricity service. Providers of customer aggregation could run local energy markets, create demand response-based incentive programs, and perform smart automatic control of their customer groups to minimize group impact.

Further, the effect of smarter scheduling of distributed energy resources should be tested. Applying this on the aggregation of solar and storage could be transformative.

\subsubsection{Cons}
\paragraph{\textbf{Peak Threshold Calculation Unfair to Solar PV?}}

As mentioned earlier, the introduction of PV could cause the total system demand to go below the peak threshold in some time slots, thus converting those time slots from peak slots to non-peak slots. However, this also shifts the peak slots to a different time, because peak slots are defined on a \textit{percentile} basis, rather than absolute. There will always be a top $x$\% set of values; it does not matter whether that range is small or large. As a result, the new system peaks would be those times when sunlight is not available. The appliance usage of a PV home is not offset when the sun is not shining. These shifted peak slots could be when the PV homes stop generating, and demand power from the grid, thus contributing to increase in the system demand. These slots are now the peak slots, and PV homes, along with all other homes, contribute to their $W$ and $V$ impact factors significantly during this time. Thus, it could lead to the situation where non-PV homes get away with highly variable or inefficient usage patterns when the sun is shining because PV homes are generating enough power to reduce the stress on the system below the system peak threshold. Essentially, some non-PV homes escape penalization due to their behavior being covered or compensated for by the PV homes.

This problem could be easily dealt with when rolling out the proposed mechanism in practice: peak thresholds could potentially be selected by DUs based on distribution feeder capacity, for the feeders on which this algorithm is being implemented. Thus, the peak thresholds would become absolute, rather than relative.

\paragraph{\textbf{Solar PV Ancillary Benefits}}
It could be argued that Solar PV is not being rewarded for the various benefits it brings to the grid or indeed its societal benefit in terms of reducing pollution. Distributed PV systems likely provide ancillary benefits such as reducing distribution system losses by generating close to the point of consumption, and in the future also might offer frequency and voltage support services through the use of smart inverters \cite{ding2016photovoltaic}.

\paragraph{\textbf{Rate Simplicity}}
Clarity and simplicity is a consideration for rate design. In this respect the volumetric rate has an advantage since customers have become accustomed to it. On the other hand it could be argued that customers are also familiar with the concept of credit scores, and paying different interest rates relative to other customers, based on their individual risk to the lender.

\paragraph{\textbf{High Grid Access Charges}}

As mentioned in \cite{BORENSTEIN20165, BATLLE2020257}, the peak-coincident charges could be too high and hence send inefficient signals to end-customers. Low income customers may not be able to respond easily to such a price signal since they have limited ability to invest in demand management technologies. Residential customers also might not have too much flexibility in their consumption schedules, due to work and school timings \cite{CHERNICK}.

\subsection{Implementation Considerations}

Our case study is a proof of concept that uses real-world data to instantiate the paradigm shift from a one-size-fits-all volumetric rate model to a customer-specific grid access fee based on cost causation. A practical implementation of a utility rate model based on this paradigm shift will require adaptation and modification to suit the existing cost-of-service regulatory framework already in place. The model would still be subject to rigorous regulatory review in the ratemaking process for the distribution utility.

In this subsection, we propose some specific practical recommendations that could be applied to make the proposed rate mechanism more implementable in the General Rate Case environment.

\subsubsection{Calculation and Periodic Update of Key Parameters}
The proposed rate mechanism employs a few key parameters which, in the study, rely on engineering judgment. When this proposed mechanism is being implemented, these parameters need to be carefully calculated because they strongly influence the final bills for all customers in the system. Critical metrics including the peak threshold ($S_\text{PeakTh}$), Strictness of Peak Demand Percentile Cut-off ($k$), Allocation \% of Total Target Revenue to Grid Impact Factors $\Pi_V$ and $\Pi_W$ all need to be defined clearly through rigorous studies. These studies to calculate key parameters can be conducted periodically as a regulatory requirement, with the engineering analysis performed by the DU, with oversight of the regulator. These parameters could remain fixed for a certain period (e.g. the duration of the rate period or year), and are revised based on periodic studies.

\subsubsection{Feeder-Specific Peak Threshold Design}
We propose calculating feeder-specific peak thresholds, so that the impact of a customer during the critical “grid stress” conditions on the feeder is captured more accurately. The utility company can conduct studies to determine the peak thresholds by considering various parameters, including typical feeder loading conditions, time of operation under increased loading conditions, and line losses, to name a few.

\subsubsection{Feeder-Specific Total Target Revenue}
We propose calculating total target revenue on a feeder-specific basis. Utility companies typically have most of the pertinent information such as capital and maintenance costs of grid infrastructure assets (such as transformers, capacitor banks and poles) already categorized by feeder. Thus, a feeder-specific approach will make the implementation of a cost-causation based rate mechanism more accurate and straightforward, compared to a system-wide approach. The responsibility to carry out such studies could easily be subsumed into the traditional roles of distribution utilities.

\subsubsection{Feeder-Specific Smart Customer Aggregation}
The objective of smart customer aggregation is to reward reductions in demand and variability impact on the grid. In other words, customers must be aggregated such that there is a physical manifestation of the desired reduction in grid impact. Therefore, we propose that each customer in an aggregated group must be on the same physical network, i.e., the same feeder.

\subsubsection{Mechanism for Implementation}\label{implementation}


Implementing this proposed rate structure is data-intensive, requiring timely data at customer-level granularity. A practical approach to implementation might be to categorize customers based on two key factors: type of customer (DER or non-DER) and the level of their impact on the grid (low, medium, and high based on the proposed metric). For each customer subgroup, the grid access fee could be the same.

\subsubsection{Determination of More Cost-Causing Grid Factors}
At present, the proposed rate mechanism involves two cost-causing Grid Factors: demand magnitude $W_i$ and demand variability $V_i$. While these two factors are impacted by several drivers of cost, such as voltage and frequency fluctuations, capacity, line losses, and age of existing equipment, this is not an exhaustive list. Further studies need to be conducted to determine more parameters that drive cost incurred by utilities when providing service to their customers.

\subsubsection{Ex-Post Review}
The existing rate mechanism is ex-ante. Utility companies can thus determine the rate charged per kWh, which is revised in every rate case. Given that the proposed rate mechanism is ex-post, and evaluates the impact of a customer based on cost-causation at the end of the observation period, traditional monthly billing needs to be adapted. This can be achieved by using historical smart meter data to evaluate customer behavior. Before the transition to the proposed rate model, customers can be notified in advance (say one year), and customer education programs can be initiated to inform the customers how the new rate model works. Further, a “grid friendliness score” can be calculated and reported to customers, similar to a credit score, which will enable customers to understand how their consumption pattern impacts their bill. A moving-window based evaluation should be adopted, so that this grid friendliness score can be evaluated and updated periodically (say monthly). These recommendations would provide adequate notice as well as evaluate customers over a substantial period to ensure fairness.

\subsubsection{Other Charges}
While we have discussed a mechanism to charge customers for some grid costs based on cost-causation principles, these do not account for all costs. Other charges include residual network costs as well as policy-related costs, which can be recovered in a minimally distortive fashion \cite{MIT_UF, BORENSTEIN20165, BATLLE2020257, Pollitt}. Utilities may be allowed to charge customers for some variable costs using riders or surcharges for specific purposes, including costs of new power plants, costs of energy efficiency programs, costs of renewable energy standards, or cross-subsidies for low-income customers \cite{xcel}.
\newline
\newline
Further research is required to substantiate the details needed for implementation.

\subsection{Policy Implications}

The current volumetric rate structure has apparent drawbacks, the first being that the DU is not assured of sufficient revenues, and the second that there is effectively an unfair cross-subsidy from non-PV customers to PV customers. With declining revenues, the DU would be forced to raise the rates for everyone, so the proposed approach provides long-term stability to the DU. PV customers could face higher bills, but this might be considered justified given that the energy they contribute may not be coincident with peak demand, which is a large driver of distribution system costs. Moreover, if such customers also had optimally operating storage, their DU charges could be reduced.

With the introduction of metrics such as peak thresholds and \% allocation, the DU has far greater flexibility to modify the rate model based on the actual costs they incur, customized for their system conditions.

The regulatory framework plays a crucial role in implementing any rate design reform. For privately-owned and some publicly-owned distribution utilities, rates are regulated by the state commissions. Regulators should be careful not to favor any particular technology and rate design should be based on the actual value of energy provided by DER assets.

\section{Concluding Remarks}\label{conclusions-section}
With the increasing penetration of DER technologies, utilities are likely to face challenges associated with the current volumetric rate design. Rate design and cost allocation are imprecise, and involve both judgment and policy goals. That being said, regulators should be open to considering alternative rate designs that allocate costs that are more directly aligned with the drivers of that cost. 

The process of establishing cost allocation and rate design methods in the real world is a complex technical and policy issue. This paper aims at providing conceptual clarity and a methodological, smart meter data-driven approach to a fair and sustainable rate structure that would provide clearer signals to different end users based on their impact on distribution grid operations. This impact is quantified by a "grid friendliness score", which calculates the customers' normalized correlation of their consumption pattern with the net power demand of the distribution system.

Future work will investigate some key questions: further insight needs to be gained on how to calculate the actual Total Target Revenue, such that it recovers the costs incurred by utilities under different system conditions. In addition to magnitude and variability of consumption, several factors may contribute to costs, such as line losses, locational congestion, and ageing of equipment. Our proposed framework allows for the addition of such contributors as individual parameters.


\subsection*{Acknowledgments}
This work is supported in part by Power Systems Engineering Research Center, and in part by National Science Foundation Grants OAC-1636772 and ECCS-1839616.

The authors are grateful to Dr. Leigh Tesfatsion and Editor-in-Chief Dr. Janice Beecher for comments that helped improve the manuscript.

\bibliographystyle{cas-model2-names}

\bibliography{cas-refs}

\end{document}